\pdfoutput=1

\documentclass[aps,pre,reprint,superscriptaddress]{revtex4-2}

\usepackage[caption=false]{subfig}
\usepackage{graphicx}
\usepackage{amsmath}
\usepackage[normalem]{ulem}
\usepackage{siunitx}
\usepackage{svg}
\usepackage{amssymb}
\usepackage{mathtools}
\usepackage{hyperref}

\newcommand*\diff{\mathop{}\!\mathrm{d}}

\newcommand{\figref}[1]{Fig.~\ref{#1}}

\newcommand{\Eqref}[1]{Eq.~\eqref{#1}}
\newcommand{\Eqsref}[1]{Eqs.~\eqref{#1}}

\newcommand{\LambdaA}{\Lambda_\mathrm{a}}
\newcommand{\LambdaP}{\Lambda_\mathrm{p}}

\newcommand{\la}[1]{l_{\mathrm{a}, #1}}
\newcommand{\lp}[1]{l_{\mathrm{p}, #1}}
\newcommand{\kBT}{k_\mathrm{B}T}

\begin{document}

\title{Size control guidelines for chemically active droplets}

\author{Guido L. A. Kusters}
\email[]{guido.kusters@ds.mpg.de}
\affiliation{Max Planck Institute for Dynamics and Self-Organization, Göttingen, Germany}

\author{David Zwicker}
\affiliation{Max Planck Institute for Dynamics and Self-Organization, Göttingen, Germany}

\date{\today}

\begin{abstract}
    Biological cells and synthetic analogues use liquid-liquid phase separation to dynamically compartmentalize their environment for various applications. In many cases, multiple droplets need to coexist, and their size needs to be controlled, which is challenging because large droplets tend to grow at the expense of smaller ones. Chemical reactions can, in principle, control droplet sizes, but there are no clear guidelines on how to robustly achieve size control. To provide guidelines, we consider a binary fluid model driven out of equilibrium by chemical reactions. We reveal two different classes of size-controlled droplets, depending on the ratio of droplet radius to the reaction-diffusion length. Moreover, we determine parameter regimes in which droplets become small. Taken together, our theory allows us to separately predict the chemical reactions necessary for maintaining droplets of a given class or size.
\end{abstract}

\maketitle

\section{Introduction}\label{sec:introduction}

Many biological and synthetic systems rely on liquid-liquid phase separation for spatial organization.
In particular, biological cells create liquid-like compartments that can regulate chemical reactions \cite{hyman2014liquid,banani2017biomolecular,alberti2017wisdom,berry2018physical,lasker2022material}, buffer protein concentrations \cite{klosin2020phase,deviri2021physical,zechner2025concentration,Monchaux2025buffering}, and respond to external stresses \cite{brangwynne2009germline,wheeler2016distinct,yoo2022chaperones}.
In a similar vein, synthetic systems have been engineered to form hierarchical multiphase structures \cite{ge2009vivo,dzuricky2020novo,tomares2022repurposing,liao2024charge}, create micro-reactors that concentrate enzymes or catalysts \cite{koga2011peptide,saha2019condensates,kuffner2020acceleration,peeples2021mechanistic,smokers2024droplets}, program droplet (dis)assembly \cite{deng2020programmable,donau2023chemistry}, and sense environmental changes \cite{lv2015photocatalytic,jambon2024phase}.
In many of these applications, droplet size needs to be controlled, which remains a central design challenge.

One approach to regulating droplet size is through the use of chemical reactions.
In the absence of reactions, droplets generally coarsen via Ostwald ripening \cite{ostwald1897studien,lifshitz1961kinetics,wagner1961theorie,Voorhees1985}, where large droplets grow at the expense of smaller ones and the system converges to a single minority-phase droplet at equilibrium.
The size of such a droplet is dictated by the coexisting concentrations and the total amount of material, implying it cannot be tuned independently.
Driven chemical reactions, however, can arrest or even reverse coarsening to enable size control.
To describe this coupling, many initial works used linear rate equations \cite{zwicker2015suppression,wurtz2018chemical,weber2019physics}, but more recent works have started to incorporate reaction schemes that explicitly obey the underlying thermodynamic constraints \cite{kirschbaum2021controlling,zwickerPhysicsDropletRegulation2025,bauermann2025theory}.
Although recent experimental advances indicate that such control can indeed be achieved \cite{dai2023engineering,sastre2025size,ura2025polymer}, clear design guidelines for realizing stable, size-controlled droplets remain elusive.

In this paper, we address the question of droplet size control using a binary fluid model that is driven out of equilibrium by chemical reactions. 
The qualitative behavior of such a droplet is determined by its size as compared to the reaction-diffusion length---the typical distance a particle diffuses before undergoing a reaction---which introduces distinct classes of ``small" and ``large" droplets.
Alternatively, one may approach size control from the perspective of geometric constraints, and classify droplets as small or large compared to the length scale imposed by such constraints.
Here, we show that the above views of droplet size control are not in general compatible, and we provide design guidelines for both.

\section{Continuum model of chemically active droplets}

\begin{figure}
	\centering{
            \includegraphics[width=0.475\textwidth]{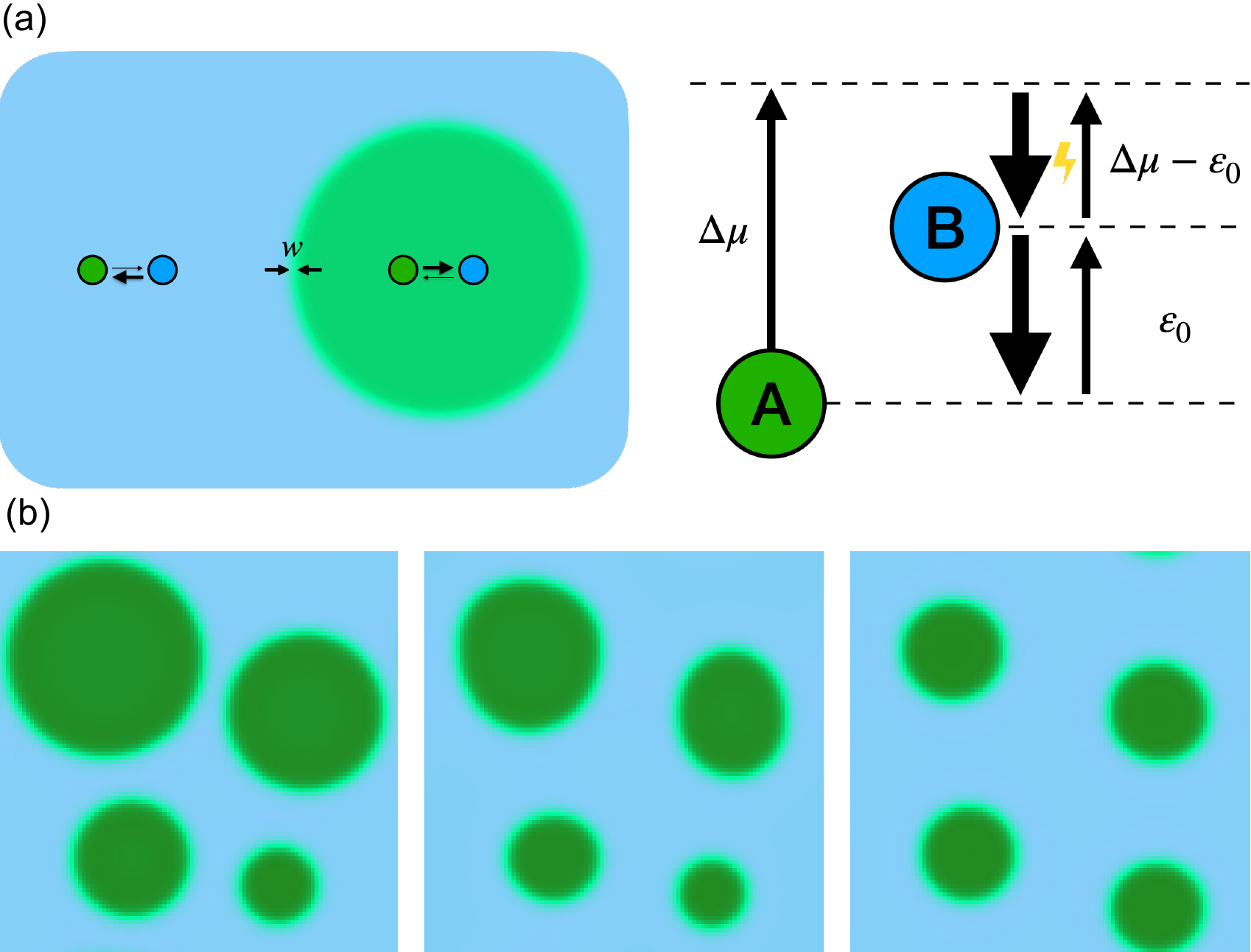}
	}
	\caption{\textbf{Conceptual model.} (a) Schematic representation of the model, in which components A and B segregate from each other and inter-convert through chemical reactions.
    The internal energy difference between the components, $\varepsilon_0$, drives a passive reaction, while an active reaction is facilitated by an additional input of external energy, $\Delta\mu$.
    Concentration-dependent reaction rates can lead to differential production in the two phases.
    (b) Concentration field $\varphi(\vec r)$ at three time points illustrating effective droplet size control.
    Simulations of \Eqsref{eq:continuity}-\eqref{eq:TST} were done on a two-dimensional periodic grid of size $100w\times100w$ and resolution $0.5w$ using \texttt{py-pde}~\cite{Zwicker2020} 
    for model parameters $f\left(\varphi\right)=\frac{B}{2}\varphi^2(1-\varphi)^2-0.05\,B\varphi$, $\Delta\mu=0.1\,B\nu$, $\Lambda_p(\varphi)=0.07[1-\tanh(\frac{\varphi-0.5}{0.01})]M$, $\Lambda_a(\varphi)=0.09[1+\tanh(\frac{\varphi-0.5}{0.01})]M$.
    }
	\label{fig:sketch}
\end{figure}

We consider an incompressible fluid comprising two components A and B that phase separate from each other.
The volume fraction field $\varphi(\vec{r}, t)$ of component~A fully describes the system since the fraction of B is $1-\varphi$.
The dynamical behavior of the fluid follows the continuity equation
\begin{equation}\label{eq:continuity}
        \frac{\partial\varphi}{\partial t}+\vec{\nabla}\cdot\vec{J}=s
        \;,
\end{equation}
where $\vec J$ is a spatial flux and the source term $s$ accounts for the conversion between A and B.
For the model to be thermodynamically consistent, both fluxes must be driven by the exchange chemical potential
\begin{equation}\label{eq:chem pot}
        \mu=\nu\frac{\delta}{\delta\varphi}\int\left[f\left(\varphi\right)+\frac{\kappa}{2}\lvert\nabla\varphi\rvert^2\right]\diff V
        \;,
\end{equation}
where $\nu$ denotes a molecular volume scale, $f\left(\varphi\right)$ the free energy density, and $\kappa$ a square-gradient coefficient that penalizes spatial gradients in composition \cite{cahn1961spinodal}.

Without reactions ($s=0$), a suitable choice of $f\left(\varphi\right)$ leads to phase separation (\figref{fig:sketch}).
The equilibrium state is then characterized by the formation of ``enriched" droplets, suspended in a ``depleted" phase.
In the thermodynamic limit, these phases have well-defined equilibrium compositions $\varphi^{(0)}_\mathrm{in}$ and $\varphi^{(0)}_\mathrm{out}$, respectively, and are separated by an interface with surface tension $\sigma$ and width $w=\sqrt{\kappa/B}$, where $B$ is the characteristic energy-density scale of $f\left(\varphi\right)$.
In this passive system, surface tension drives coarsening~\cite{Voorhees1985}, which implies that equilibria contain at most a single droplet.
In finite systems, this droplet is spherical, characterized by a radius $R$.
The coexisting compositions associated with this finite system are slightly elevated due to the surface tension effects, $\varphi^\mathrm{eq}_\mathrm{in}\approx\varphi^{(0)}_\mathrm{in}$ and $\varphi^\mathrm{eq}_\mathrm{out}\approx\varphi^{(0)}_\mathrm{out}[1+2\sigma\nu/(R\varphi^{(0)}_\mathrm{in}\kBT)]$, where $\kBT$ denotes the thermal energy~\cite{thomson18724,weber2019physics,zwickerPhysicsDropletRegulation2025}.
The physics of droplet formation is thus well-understood in passive systems, but activity modifies the picture.

To study the kinetics of active systems, we focus on linear non-equilibrium thermodynamics, where the spatial flux $\vec{J}=-M(\varphi)\vec{\nabla}\mu$ is driven by gradients in $\mu$ proportional to the diffusive mobility $M(\varphi)$ \cite{onsager1931reciprocal,dover1984}.
In contrast, the reaction flux $s$ is directly driven by $\mu$,
\begin{equation}\label{eq:TST}
    s=-\LambdaP\left(\varphi\right)\mu-\LambdaA\left(\varphi\right)\left(\mu+\Delta\mu\right)
    \;.
\end{equation}
The two terms in \Eqref{eq:TST} respectively describe a \emph{passive} reaction, whose conversion rate is proportional to $\mu$, and an \emph{active} reaction driven by an additional external energy input $\Delta\mu$ (\figref{fig:sketch}(a)). 
The associated fluxes are proportional to the respective reaction mobilities $\LambdaP$ and $\LambdaA$, which can depend on composition $\varphi$.
For instance, biological reactions are often accelerated by enzymes, which might accumulate in the droplet.
Taken together, Eqs.~\eqref{eq:continuity}-\eqref{eq:TST} can arrest or even reverse Ostwald ripening and give rise to droplet size control (\figref{fig:sketch}(b)), in contrast to an equilibrium system. 

\section{Thin-interface approximation}

To analyze the dynamics of chemically active droplets in detail, we assume that reactions are weak, such that equilibrium phase coexistence holds approximately.
More precisely, we assume that reactions do not significantly affect surface tension, interface width, and coexisting compositions at the droplet interface, so we can use the respective expressions for equilibrium interfaces.
Droplets are then typically spherical, so we can parameterize them with their radius~$R$.
Assuming relatively large droplets ($R\gg w$), we then consider a thin interface and only account for reactions in the adjacent bulk regions.
The dynamics of $R$ are governed by the balance of diffusive fluxes at the interface,
\begin{equation}
    \label{eq:flux_balance}
    \dot{R} = \frac{1}{\Delta\varphi}\left(-M_\mathrm{in}\frac{\partial \mu_\text{in}}{\partial r}\bigg\rvert_{R}+M_\mathrm{out}\frac{\partial \mu_\text{out}}{\partial r}\bigg\rvert_{R}\right)
    \;,
\end{equation}
where the subscripts refer to the droplet phase (``in") and the surrounding dilute phase (``out"), respectively, and $\Delta\varphi=\varphi_\mathrm{in}^\mathrm{eq}-\varphi_\mathrm{out}^\mathrm{eq}$.
Here, we have approximated the diffusive mobilities at either side of the thin interface, $M_\mathrm{in}=M(\varphi^{(0)}_\mathrm{in})$ and $M_\mathrm{out}=M(\varphi^{(0)}_\mathrm{out})$, by their values at the equilibrium compositions.
The right-hand side of \Eqref{eq:flux_balance} depends on the chemical potential profiles~$\mu_\alpha(\vec r)$ in both phases.
In the case of weak reactions, the variation of $\mu_\alpha(\vec r)$ is limited, so we determine it by expanding \Eqref{eq:continuity} in the two phases $\alpha=\mathrm{in}$ and $\alpha=\mathrm{out}$,
\begin{equation}\label{eq:piecewise_Eqn}
      \frac{\partial \varphi_\alpha}{\partial t} = M_\alpha\left(\nabla^2\mu_\alpha-\frac{1}{\lp{\alpha}^2}\mu_\alpha-\frac{1}{\la{\alpha}^2}\left(\mu_\alpha+\Delta\mu\right)\right)
      \;,
\end{equation}
where we assume that the aforementioned diffusive mobilities at either side of the thin interface, $M_\alpha$, hold approximately throughout their respective phases.
A similar approximation for reactive mobilities allows us to introduce reaction-diffusion lengths $\lp{\alpha}=[M_\alpha/\LambdaP(\varphi^{(0)}_\alpha)]^{1/2}$ and $\la{\alpha}=[M_\alpha/\LambdaA(\varphi^{(0)}_\alpha)]^{1/2}$ for the passive and active reaction, respectively.
\Eqref{eq:piecewise_Eqn} is subject to a no-flux condition at the system's exterior boundary, which we take to be at infinity, implying sufficiently large systems. 
In contrast, equilibrium phase coexistence holds at the interface between the two phases, implying $\varphi_\mathrm{in}=\varphi_\mathrm{in}^\mathrm{eq}$, $\varphi_\mathrm{out}=\varphi_\mathrm{out}^\mathrm{eq}$, and $\mu_\mathrm{in}=\mu_\mathrm{out}=(-\varepsilon_0+2\sigma\nu R^{-1})/\Delta\varphi$ at the interface~\cite{thomson18724,bray1994theory,weber2019physics,zwickerPhysicsDropletRegulation2025}, where $\varepsilon_0=-\nu [f(\varphi^\mathrm{eq}_\mathrm{in})-f(\varphi^\mathrm{eq}_\mathrm{out})]$ is the free energy difference per molecule between the two phases.
The combination of \Eqsref{eq:flux_balance} and \eqref{eq:piecewise_Eqn} fully determine the interface kinetics of an isolated active droplet.

\begin{figure}
    	\centering{
            \includegraphics[width=0.475\textwidth]{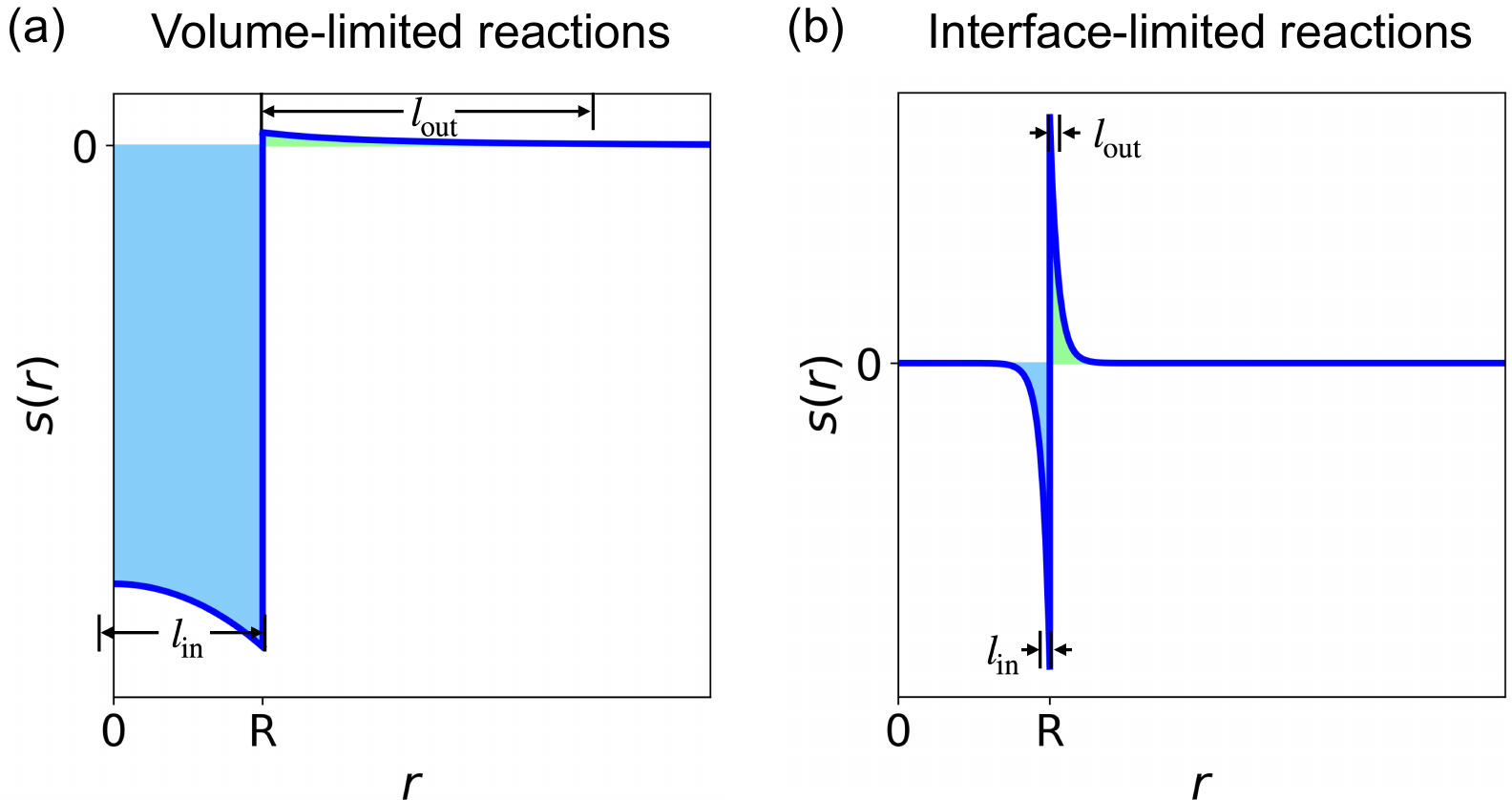}
	}
	\caption{
    \textbf{Volume-limited and interface-limited reactions.}
    Quasi-static reaction flux $s$ as a function of the radial coordinate $r$ for different values of $l_\mathrm{in}$. The shaded blue and teal areas correspond to degradation and production of droplet material, respectively.
    (a) Model parameters are
    $\sigma=0.15\,Bw$,
    $\Delta\varphi=1$,
    $\varepsilon_0=0.05\,B\nu$,
    $\varepsilon_\mathrm{in}=0.05\,B\nu$,
    $\varepsilon_\mathrm{out}=-0.05\,B\nu$,
    $l_\mathrm{in}=10^4\,w$,
    $l_\mathrm{out}=1.11\cdot10^4\,w$, and
    $R=2\cdot 10^5\,w$.
    (b) Model parameters are
    $\sigma=0.15\,Bw$,
    $\Delta\varphi=1$,
    $\varepsilon_0=0.01\,B\nu$,
    $\varepsilon_\mathrm{in}=0.09\,B\nu$,
    $\varepsilon_\mathrm{out}=-0.01\,B\nu$,
    $l_\mathrm{in}=500\,w$,
    $l_\mathrm{out}=1000\,w$, and
    $R=455\,w$.
    }
	\label{fig:s_r}
\end{figure}

To see when chemically active droplets are stable, we next solve \Eqref{eq:piecewise_Eqn} for the chemical potential profiles $\mu_\alpha\left(r\right)$, where $r$ denotes the radial coordinate centered at the droplet center.
Since the time scale on which the composition field $\varphi$ relaxes is typically much faster than the time scale on which the droplet interface moves~\citep{bray1994theory}, we solve \Eqref{eq:piecewise_Eqn} in stationary state,
\begin{subequations}
\label{eq:piecewise}
\begin{align}
    \mu_\mathrm{in}(r) &= 
    \frac{\varepsilon_\mathrm{in}+\frac{2\sigma\nu}{R}}{\Delta\varphi}\cdot\frac{\sinh\!\left(\frac{r}{l_\mathrm{in}}\right)}{\sinh\!\left(\frac{R}{l_\mathrm{in}}\right)}\frac{R}{r}
    -\frac{\varepsilon_\mathrm{in}+\varepsilon_0}{\Delta\varphi}
    , 
    &r&\leq R,
\\
    \mu_\mathrm{out}(r) &=
    \frac{\varepsilon_\mathrm{out}+\frac{2\sigma\nu}{R}}{\Delta\varphi}\cdot\frac{e^{-\frac{r}{l_\mathrm{out}}}}{e^{-\frac{R}{l_\mathrm{out}}}}\frac{R}{r}
    -\frac{\varepsilon_\mathrm{out}+\varepsilon_0}{\Delta\varphi}
    , &r&>R,
\end{align}
\end{subequations}
introducing the effective parameters
\begin{align}
\label{eq:eff_params}
    l_\alpha&= \Bigl(\la{\alpha}^{-2}+\lp{\alpha}^{-2}\Bigr)^{-\frac12},
&
    \varepsilon_\alpha&= \frac{\Delta\mu\Delta\varphi}{1+\la{\alpha}^2\lp{\alpha}^{-2}}-\varepsilon_0 \;,
\end{align}
for $\alpha = \text{in}, \text{out}$.
To interpret these parameters, we substitute \Eqsref{eq:piecewise} into \Eqref{eq:TST} to obtain the local production of droplet material (\figref{fig:s_r}).
Looking at both phases separately, we find that the magnitude of the net turnover rate $|s|$ is greatest at the droplet interface, where its value is given by $s_\alpha=-M_\alpha(\varepsilon_\alpha+2\sigma\nu R^{-1})/(l_\alpha^2\Delta\varphi)$ in the respective phases, and decays further away from it.
This identity suggests that $\varepsilon_\alpha$ contributes to the magnitude of the conversion rate and, perhaps more importantly, determines its sign, i.e., whether droplet material is locally produced ($s>0$) or degraded ($s<0$).
\Eqsref{eq:piecewise} show that the conversion rate falls off with characteristic length scales $l_\alpha$ away from the interface.
This suggests chemical reactions inside the droplet may either occur throughout the entire droplet ($R\ll l_\mathrm{in}$, \figref{fig:s_r}(a)) or be concentrated at the interface ($R\gg l_\mathrm{in}$, \figref{fig:s_r}(b)).

To analyze the detailed dynamics of the droplet radius~$R$, we insert \Eqsref{eq:piecewise} into \Eqref{eq:flux_balance} to obtain
\begin{align}
    \dot{R}&=-\frac{1}{R\Delta\varphi^2}\biggl[M_\mathrm{out}\left(\varepsilon_\mathrm{out}+\frac{2\sigma\nu}{R}\right)\left(1+\frac{R}{l_\mathrm{out}}\right)
\notag\\
    &\quad+M_\mathrm{in}\left(\varepsilon_\mathrm{in}+\frac{2\sigma\nu}{R}\right)\left(\frac{R}{l_\mathrm{in}}\coth\!\left({\frac{R}{l_\mathrm{in}}}\right)-1\right)\biggr]
    \;.
\label{eq:Req}
\end{align}
The first term in the square brackets stems from the chemical reactions---and corresponding diffusive fluxes---outside the droplet, whereas the second term stems from reactions inside the droplet.
The relative importance of these terms is governed by the diffusive mobilities $M_\alpha$, the energy scales $\varepsilon_\alpha$, and the reaction-diffusion lengths~$l_\alpha$.
However, in the simple case without surface tension ($\sigma=0$), only the product $M_\alpha \epsilon_\alpha$ appears, indicating that these two parameters play similar roles.
Since surface tension effects are typically small, we thus consider $M_\mathrm{in}=M_\mathrm{out}$ in what follows.
Taken together, the key model parameters are $\varepsilon_\alpha$ and $l_\alpha$, which influence the droplet radius~$R$ as the sole dynamical degree of freedom, governed by \Eqref{eq:Req}.

\begin{figure}
        	\centering{
            \includegraphics[width=0.475\textwidth]{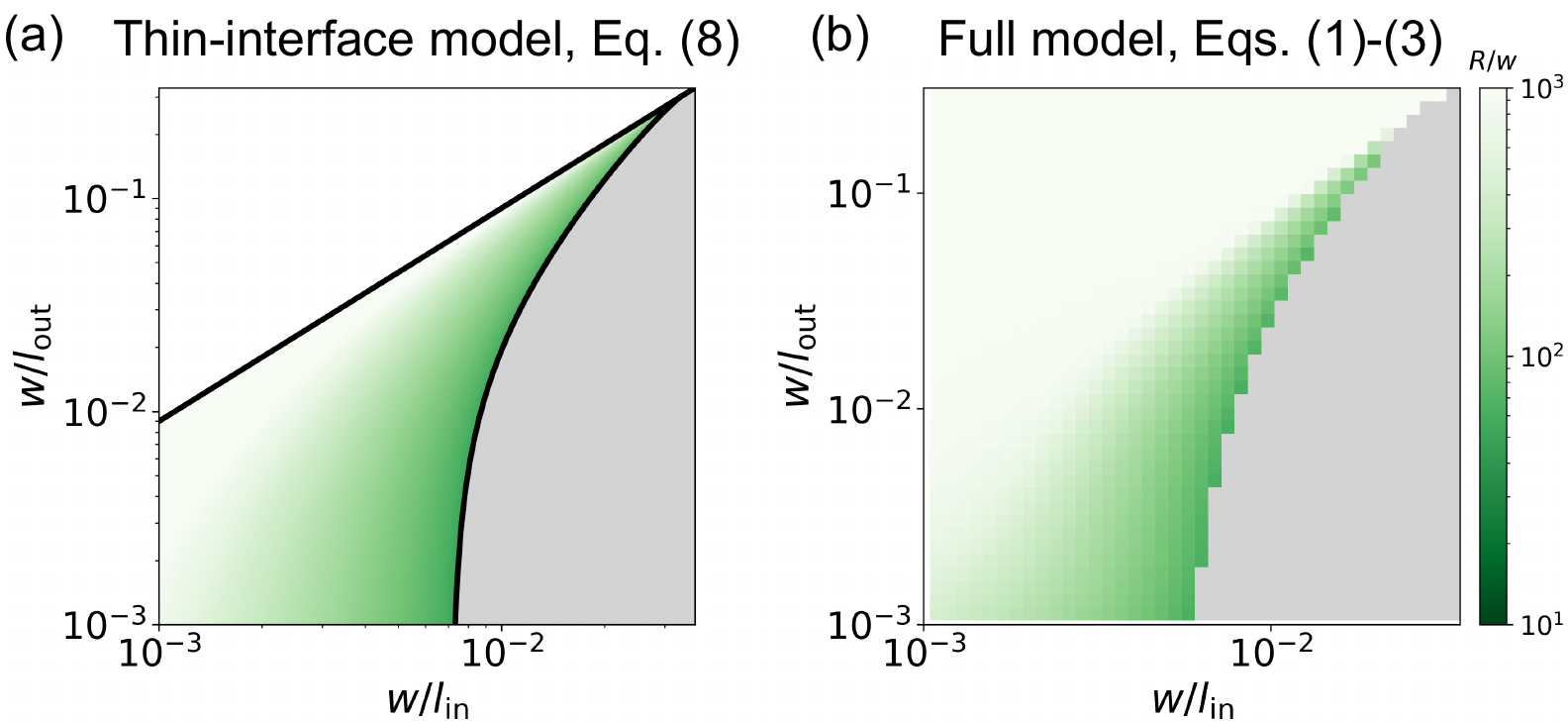}
	}
    \caption{
    \textbf{Thin-interface model captures stationary state radii.}
    Steady state droplet radius~$R$ as a function of the scaled reaction-diffusion lengths $w/l_\mathrm{in}$ and $w/l_\mathrm{out}$.
    The system is homogeneously enriched in component A (white region), homogeneously enriched in component B (gray region), or exhibits an isolated droplet whose radius is indicated by color.
    (a) Data based on the stable fixed points of Eq. \eqref{eq:Req} for model parameters $\varepsilon_\mathrm{in}=0.09B\nu$, $\varepsilon_\mathrm{out}=-0.01B\nu$, $\sigma=0.15B$, and $M_\mathrm{in}=M_\mathrm{out}=M$.
    (b) Data based on simulations of Eqs.~\eqref{eq:continuity}-\eqref{eq:TST} on a three-dimensional, spherically symmetric grid of radius $1000w$ and resolution $0.5w$ using \texttt{py-pde}~\cite{Zwicker2020} for the equivalent model parameters $f\left(\varphi\right)=\frac{B}{2}\varphi^2(1-\varphi)^2-0.01B\varphi$, $\Delta\mu=0.1\,B\nu$, $\Lambda_p(\varphi)=l_\mathrm{out}^{-2}[1-\tanh(\frac{\varphi-0.5}{0.01})]M$, $\Lambda_a(\varphi)=l_\mathrm{in}^{-2}[1+\tanh(\frac{\varphi-0.5}{0.01})]M$.
    }\label{fig:phase_comp}
\end{figure}

To check how well our approximate model captures the dynamics of the full system, we compare the stable fixed points of \Eqref{eq:Req} with the final states observed from continuum simulations of \Eqref{eq:continuity}.
Fig.~\ref{fig:phase_comp} shows that both methods agree very well, motivating us to use \Eqref{eq:Req} as a predictive tool for understanding the behavior of chemically active droplets.

To see under which conditions droplets can be stable, we next analyze the steady states of \Eqref{eq:Req} in detail.
In the steady state, the reactions in the system balance globally, such that on average no net conversion between components persists.
Since the signs of the energetic prefactors $\varepsilon_\alpha+2\sigma\nu/R$ dictate the direction of conversion (positive if droplet material is degraded and negative if it is produced), they must be opposite in the two regions to enable a non-trivial steady state.
If we wish such a steady state to hold for arbitrarily large droplets ($2\sigma\nu/R\approx0$), this leaves us with two distinct choices for $\varepsilon_\alpha$: $\varepsilon_\mathrm{in}>0$, $\varepsilon_\mathrm{out}<0$ for \emph{externally} maintained droplets (i.e., the droplet component A is produced outside the droplet and degraded inside the droplet) and $\varepsilon_\mathrm{in}<0$, $\varepsilon_\mathrm{out}>0$ for \emph{internally} maintained droplets (i.e., the droplet component A is produced inside the droplet and degraded outside the droplet).
A straightforward stability analysis reveals that only the former class admits \emph{stable} steady-state solutions to \Eqref{eq:Req}~\citep{weber2019physics,zwicker2022intertwined,zwickerPhysicsDropletRegulation2025}.
We thus focus on externally maintained droplets ($\varepsilon_\mathrm{in}>0$, $\varepsilon_\mathrm{out}<0$) in the rest of this paper.

\section{Volume- and interface-limited reactions}

We start by analyzing the qualitative behavior of externally maintained droplets.
To this end, we single out the internal reaction-diffusion length $l_\mathrm{in}$ as the relevant length scale since it sets the distance over which the net conversion rate $s$ extends into the droplet (\figref{fig:s_r}).
If $R\ll l_\mathrm{in}$, the droplet is much smaller than the typical distance molecules diffuse before undergoing a chemical conversion (\figref{fig:s_r}(a)).
Consequently, we call such droplets ``volume-limited", as chemical conversions occur more or less homogeneously throughout the volume of the droplet.
Conversely, if $R\gg l_\mathrm{in}$, molecules that enter the droplet typically undergo a chemical conversion close to the interface, so we label such droplets ``interface-limited" (\figref{fig:s_r}(b)).
The ratio $R/l_\mathrm{in}$ thus provides a natural interpretation of what it means for a droplet to be ``small" or ``large," with distinct qualitative behavior for each.

\begin{figure}
    \centering{
            \includegraphics[width=0.475\textwidth]{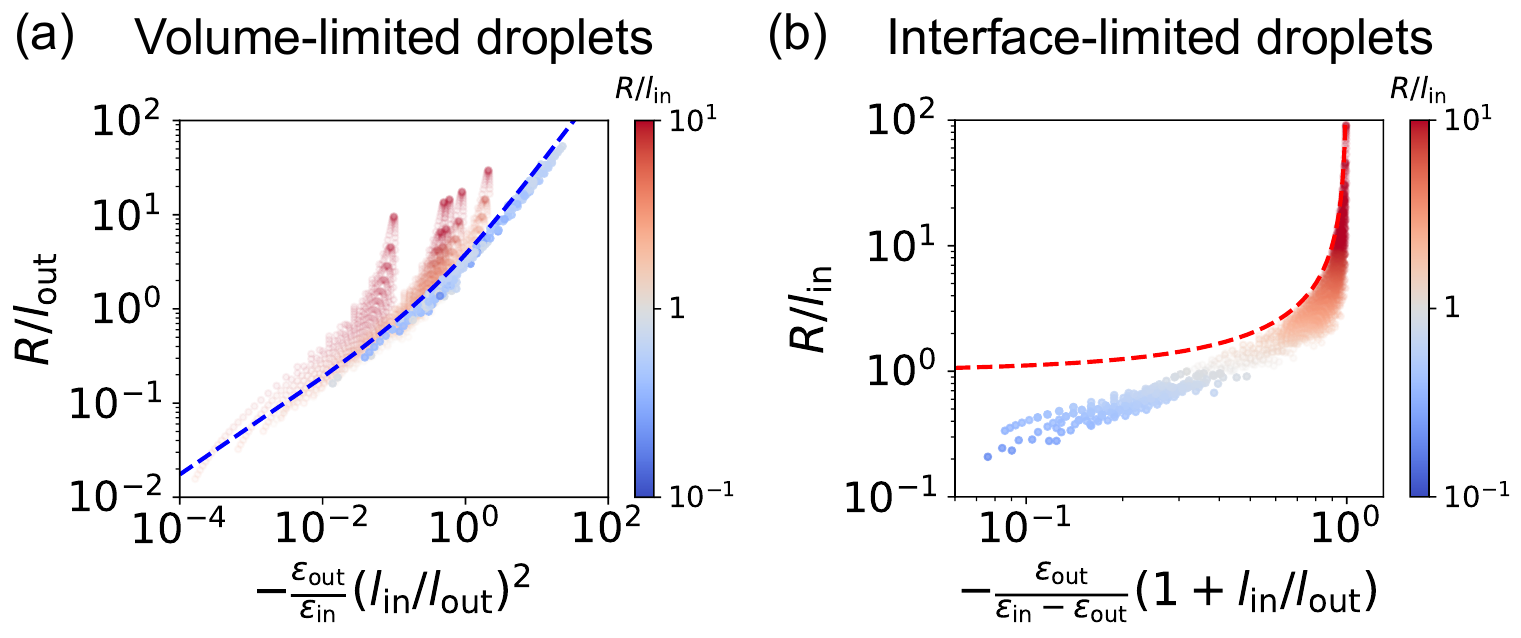}
	}
    \caption{
    \textbf{Volume-limited and interface-limited droplets obey scaling laws.}
    (a) Droplet radius~$R$ normalized by the external reaction-diffusion length $l_\mathrm{out}$ as a function of the control parameter expected for volume-limited droplets in comparison to the scaling law given by \Eqref{eq:limiting_cases} (dashed blue line).
    (b) $R$ normalized to internal reaction-diffusion length $l_\mathrm{in}$ as a function of the control parameter expected for interface-limited droplets in comparison to the scaling law given by \Eqref{eq:limiting_cases2} (dashed red line).
    (a, b)
    The colored data points represent stable fixed points of \Eqref{eq:Req} for a range of parameter values. The color-coding is prescribed by the colorbar.
    Model parameters are $\sigma=0.15\,Bw$.}\label{fig:scaling_laws}
\end{figure}

We infer the behavior of volume- and interface-limited droplets by expanding \Eqref{eq:Req} for $R/l_\mathrm{in}\ll1$ and $R/l_\mathrm{in}\gg1$, respectively, and solving for the stationary state.
To build intuition, we first focus on droplets for which $R\gg 2\sigma\nu/\textrm{min}\left(\varepsilon_\mathrm{in},|\varepsilon_\mathrm{out}|\right)$, such that surface tension effects can be neglected.
This is often a reasonable approximation if $R,l_\mathrm{in}\gg w$, since the length scale $2\sigma\nu/\textrm{min}\left(\varepsilon_\mathrm{in},|\varepsilon_\mathrm{out}|\right)$ is in many typical cases comparable to the molecular length scale $\nu^{1/3}$~\cite{pitaevskii2012physical,weber2019physics}.
In this limit, simple scaling laws emerge:
\begin{subequations}
\label{eqn:scaling_laws}
\begin{align}
    \frac{R}{l_\mathrm{out}}&\approx-\frac{3}{2}\frac{\varepsilon_\mathrm{out}}{\varepsilon_\mathrm{in}}\left(\frac{l_\mathrm{in}}{l_\mathrm{out}}\right)^2\left(1+\sqrt{1-\frac{4}{3}\frac{\varepsilon_\mathrm{in}}{\varepsilon_\mathrm{out}}\left(\frac{l_\mathrm{out}}{l_\mathrm{in}}\right)^2}\right),
\notag\\
    &\hspace{15em} R/l_\mathrm{in}\ll1,
    \label{eq:limiting_cases}
\\[10pt]
    \frac{R}{l_\mathrm{in}}&\approx\frac{1}{1+\dfrac{\varepsilon_\mathrm{out}}{\varepsilon_\mathrm{in}-\varepsilon_\mathrm{out}} \left(1+\dfrac{l_\mathrm{in}}{l_\mathrm{out}}\right)}, \qquad R/l_\mathrm{in}\gg1
    \;,
    \label{eq:limiting_cases2}    
\end{align}
\end{subequations}
where $\varepsilon_\mathrm{in}>0$ and $\varepsilon_\mathrm{out}<0$ must hold for the stationary solutions to be stable.
In either limit, 
the droplet radius~$R$ is governed by only two effective parameters, which suggests that the dependence on the four basic parameters ($\varepsilon_\mathrm{in}$, $\varepsilon_\mathrm{out}$, $l_\mathrm{in}$, and $l_\mathrm{out}$) can be drastically reduced.
Indeed, numerically determined stationary state radii are well approximated by the scaling laws given by \Eqsref{eqn:scaling_laws} for volume-limited (blue data in \figref{fig:scaling_laws}(a)) and interface-limited (red data in \figref{fig:scaling_laws}(b)) droplets as a function of the reduced parameters.

\begin{figure}
	\centering{
            \includegraphics[width=0.475\textwidth]{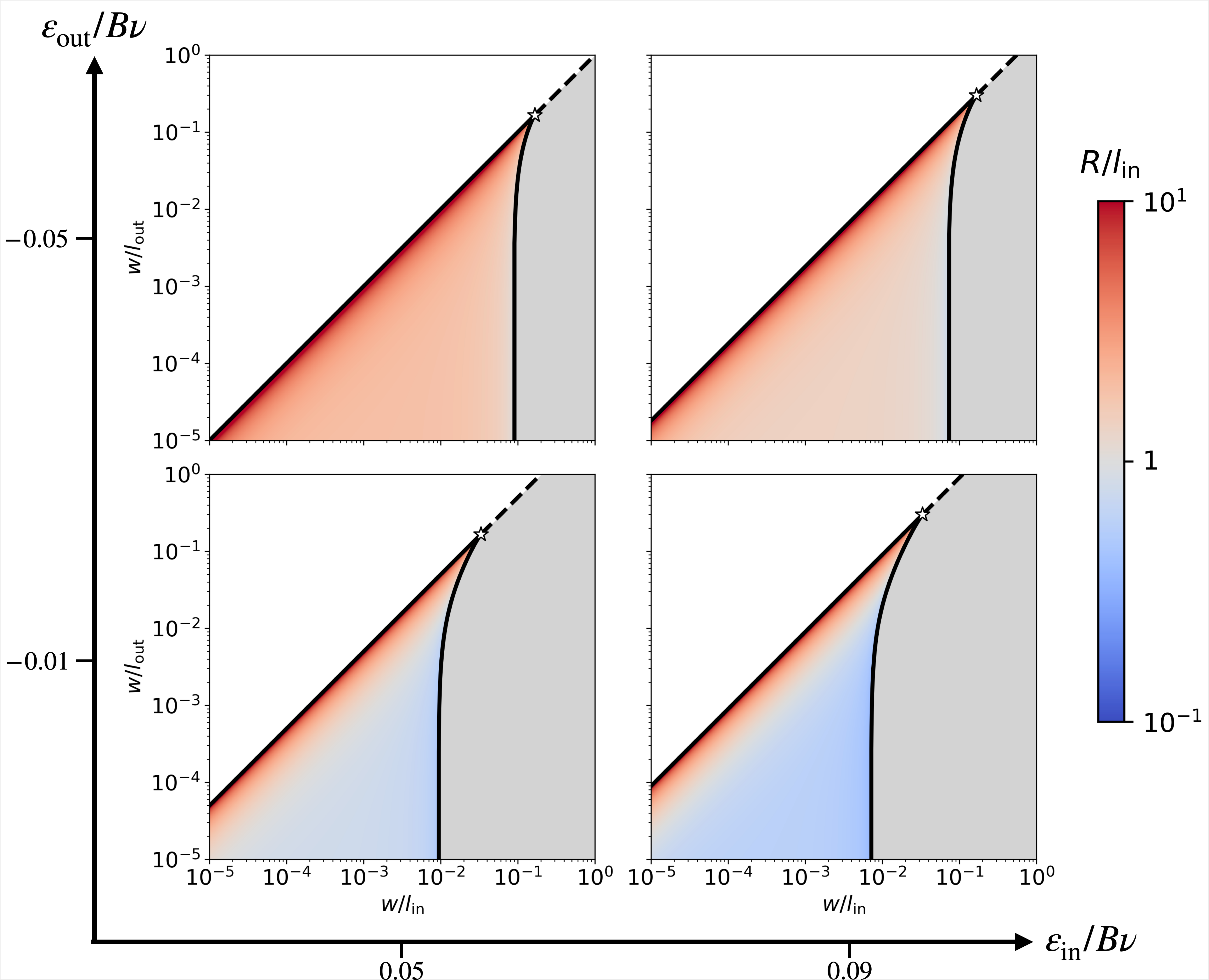}
	}
	\caption{
    \textbf{Parameter regions leading to volume- vs. interface-limited droplets.}
    Relative steady state droplet radius~$R/l_\mathrm{in}$ as a function of the scaled reaction-diffusion lengths $w/l_\mathrm{in}$ and $w/l_\mathrm{out}$. Each panel corresponds to a different combination of energy scales $\varepsilon_\mathrm{in}/B\nu$ and $\varepsilon_\mathrm{out}/B\nu$.
    The system is homogeneously enriched in component A (white region), homogeneously enriched in component B (gray region), or exhibits an isolated droplet whose radius is indicated by color.
    Model parameters are $\sigma=0.15\,Bw$.
    }
	\label{fig:rel_phase_log}
\end{figure}

To further probe under which conditions steady-state droplets of either regime can be expected, we investigate the stable fixed points of \Eqref{eq:Req} in more detail.
We first ask which combinations of the model parameters $l_\mathrm{in}$, $l_\mathrm{out}$, $\varepsilon_\mathrm{in}$, and $\varepsilon_\mathrm{out}$ allow for steady-state droplets at all (\figref{fig:rel_phase_log}).
Our analysis reveals that, regardless of the values of $\varepsilon_\mathrm{in}$ and $\varepsilon_\mathrm{out}$, stable droplets are precluded if either $l_\mathrm{in}$ is small and $l_\mathrm{out}$ is large, or $l_\mathrm{in}$ is large and $l_\mathrm{out}$ is small. 
This is because the reaction-diffusion lengths represent the relative importance of diffusion to reactions, with smaller values corresponding to faster reactions.
As a result, the former case corresponds to a scenario in which the reactions produce droplet material much more quickly than they degrade it, such that the steady state becomes a homogeneous enriched phase (white regions in \figref{fig:rel_phase_log}).
The latter case corresponds to reactions that degrade droplet material much more quickly than they produce it, leading to a homogeneous depleted phase as the steady state (gray regions in \figref{fig:rel_phase_log}).
Consequently, steady-state droplets (colored region in \figref{fig:rel_phase_log}) demand a balance between the reaction-diffusion length scales inside and outside of the droplet.

More concretely, the structure of the state diagram shown in \figref{fig:rel_phase_log} suggests that steady-state droplets can only exist if (i) their growth is halted at large droplet sizes (violated in the white region), and (ii) they overcome the inhibiting effect of surface tension (violated in the gray region).
We quantify the region that admits steady-state droplets by determining when these conditions are met.
Mathematically, the former condition demands that $\dot{R}<0$ for large $R$.
To locate the boundary beyond which condition (i) is violated, we thus take the limit of $R\rightarrow\infty$ in \Eqref{eq:Req} and solve for $\dot{R}=0$.
Rearranging terms yields the asymptote
\begin{equation}\label{eq:asymptote}
      \frac{l_\mathrm{in}}{l_\mathrm{out}}=-\frac{\varepsilon_\mathrm{in}}{\varepsilon_\mathrm{out}}
      \;.
\end{equation}
The second condition demands that $\dot{R}>0$ for some range of $R$, since otherwise droplets of all sizes would shrink and eventually dissolve.
Consequently, we find the boundary beyond which condition (ii) is violated by demanding that the growth rate $\dot R$ peaks at $\dot{R}=0$.
The parameter region where finite droplets are viable is enclosed by these two boundaries, which terminate at a critical point.
The location of this critical point follows from expanding \Eqref{eq:Req} for large $R$ and setting the result to zero.
The leading-order contribution reproduces \Eqref{eq:asymptote},
and the sub-leading-order contribution determines whether $\dot{R}$ approaches its asymptotic value from above or below.
Requiring that also this contribution vanishes gives 
\begin{align}\label{eq:critical point}
    l_{\mathrm{out},*}&=\frac{2\sigma\nu}{\varepsilon_\mathrm{in}},
&
    l_{\mathrm{in},*}&=-\frac{2\sigma\nu}{\varepsilon_\mathrm{out}}
    \;;
\end{align}
see white star in \figref{fig:rel_phase_log}. 
In particular, the critical point given by \Eqref{eq:critical point} helps us to limit the parameters for which droplet formation is possible.

To quantify admissible parameter regimes in more detail, we next measure the size of the region that exhibits stable droplets in steady state. 
As we vary $\varepsilon_\mathrm{in}$ and $\varepsilon_\mathrm{out}$, the slope of the dashed asymptote and the location of the critical point shift, expanding or contracting this region. 
Taking the area under the asymptote given by \Eqref{eq:asymptote} up to the critical point given by \Eqref{eq:critical point} as a rough estimate of this region's size, $-\frac{1}{8}\varepsilon_\mathrm{in}\varepsilon_\mathrm{out}w^2\sigma^{-2}\nu^{-2}$, we find that the product of energy scales $\varepsilon_\mathrm{in}\varepsilon_\mathrm{out}$ plays a key role in maximizing the number of reaction-rate combinations that lead to steady-state droplets.
By symmetry of \Eqsref{eq:asymptote} and \eqref{eq:critical point}, the same holds for the product $(l_\mathrm{in}l_\mathrm{out})^{-1}$ if one wishes to instead maximize the number of energy-scale combinations that yield steady-state droplets.
We thus predict that increasing either $\varepsilon_\mathrm{in}\varepsilon_\mathrm{out}$ or $(l_\mathrm{in}l_\mathrm{out})^{-1}$ leads to more stable droplets.
These trends are robust up to system-specific upper bounds for $\varepsilon_\mathrm{in}$ and $|\varepsilon_\mathrm{out}|$, beyond which an instability can be triggered that leads to spinodal decomposition in the droplet phase or the dilute phase, respectively.

We next investigate how the model parameters can be tuned to access the regimes of volume- and interface-limited droplets, for which the ratio $R/l_\mathrm{in}$ is the determining factor.
We find that increasing either the energy scale ($|\varepsilon_\mathrm{out}|$) or rate ($1/l_\mathrm{out}$) of the reaction that produces droplet material leads to an increase in $R/l_\mathrm{in}$ (blue to red in \figref{fig:rel_phase_log}), as this biases the system to a steady state with a higher average volume fraction of droplet material.
Conversely, increasing the energy scale ($\varepsilon_\mathrm{in}$) or rate ($1/l_\mathrm{in}$) of the reaction that degrades droplet material leads to a decrease in $R/l_\mathrm{in}$ (red to blue in \figref{fig:rel_phase_log}). 
Amplifying the chemical reactions (through variation of $\varepsilon_\alpha$ and $l_\alpha$) thus has opposite effects depending on whether they occur inside or outside the droplet.

To convert our observations into actionable design guidelines, recall that the effective parameters $l_\alpha$ and $\varepsilon_\alpha$ given by \Eqref{eq:eff_params} depend on a variety of ``bare" model parameters: the relevant active and passive reaction-diffusion lengths, $l_{\mathrm{a},\alpha}$ and $l_{\mathrm{p},\alpha}$, the internal energy difference $\varepsilon_0$ between components A and B, and the chemical drive $\Delta\mu$.
Notably, our conclusions in terms of effective parameters do not translate one-to-one to these ``bare" parameters, as both $l_\alpha$ and $\varepsilon_\alpha$ depend on $l_{\mathrm{a},\alpha}$ and $l_{\mathrm{p},\alpha}$.
As a result, variation of $l_\alpha$ does not, in general, keep $\varepsilon_\alpha$ constant.
To nevertheless put our results into perspective, we consider the specific case where the chemical interconversion inside the droplet is dominated by an active reaction that degrades droplet material ($l_\mathrm{in}= \sqrt{M/\LambdaA}$, $\varepsilon_\mathrm{in}=\Delta\mu\Delta\varphi-\varepsilon_0$), whereas the chemical interconversion outside the droplet is dominated by a passive reaction that produces droplet material ($l_\mathrm{out}= \sqrt{M/\LambdaP}$, $\varepsilon_\mathrm{out}=-\varepsilon_0$). 
This introduces the passive reaction rate constant~$\LambdaP$, the internal energy difference~$\varepsilon_0$ between components A and B, the active reaction rate constant~$\LambdaA$, and the chemical drive~$\Delta\mu$ as independent and accessible control parameters. 
The former two are linked to the production of droplet material and increasing them promotes interface-limited droplets, whereas the latter two are linked to degradation and increasing them promotes volume-limited droplets.
In addition, we showed above that the ``area" of the droplet-forming region in \figref{fig:rel_phase_log} (i.e., the number of reaction-rate combinations $(l_\mathrm{in}, l_\mathrm{out})$ that admit steady-state droplets) scales with the product $|\varepsilon_\mathrm{in}\varepsilon_\mathrm{out}|$.
If we substitute into this expression the parameters applicable to our specific case, it follows that the number of accessible reaction rate constant combinations $(\LambdaP, \LambdaA)$ grows with increasing $\Delta\mu$ and reaches its maximum for $\varepsilon_0=\frac12\Delta\mu\Delta\varphi$.
By symmetry, the number of admissible energy-scale combinations $(\varepsilon_\mathrm{in}, \varepsilon_\mathrm{out})$ scales with the product $l_\mathrm{in}l_\mathrm{out}$, which in our specific case translates to the number of admissible $(\varepsilon_0, \Delta\mu)$ combinations scaling with the product of reaction rate constants $\LambdaP\LambdaA$.

Together, our findings show that the relative importance of external and internal parameters robustly determines whether volume- or interface-limited droplets form, regardless of the system specifications. 
The corresponding size scaling is well captured by \Eqsref{eqn:scaling_laws}.
In contrast, depending on the specific system, the likelihood of realizing a steady-state droplet in the first place is not necessarily improved by maximizing the available system parameters, but can exhibit intermediate optima.

\section{Droplet size control}

\begin{figure}
	\centering{
            \includegraphics[width=0.475\textwidth]{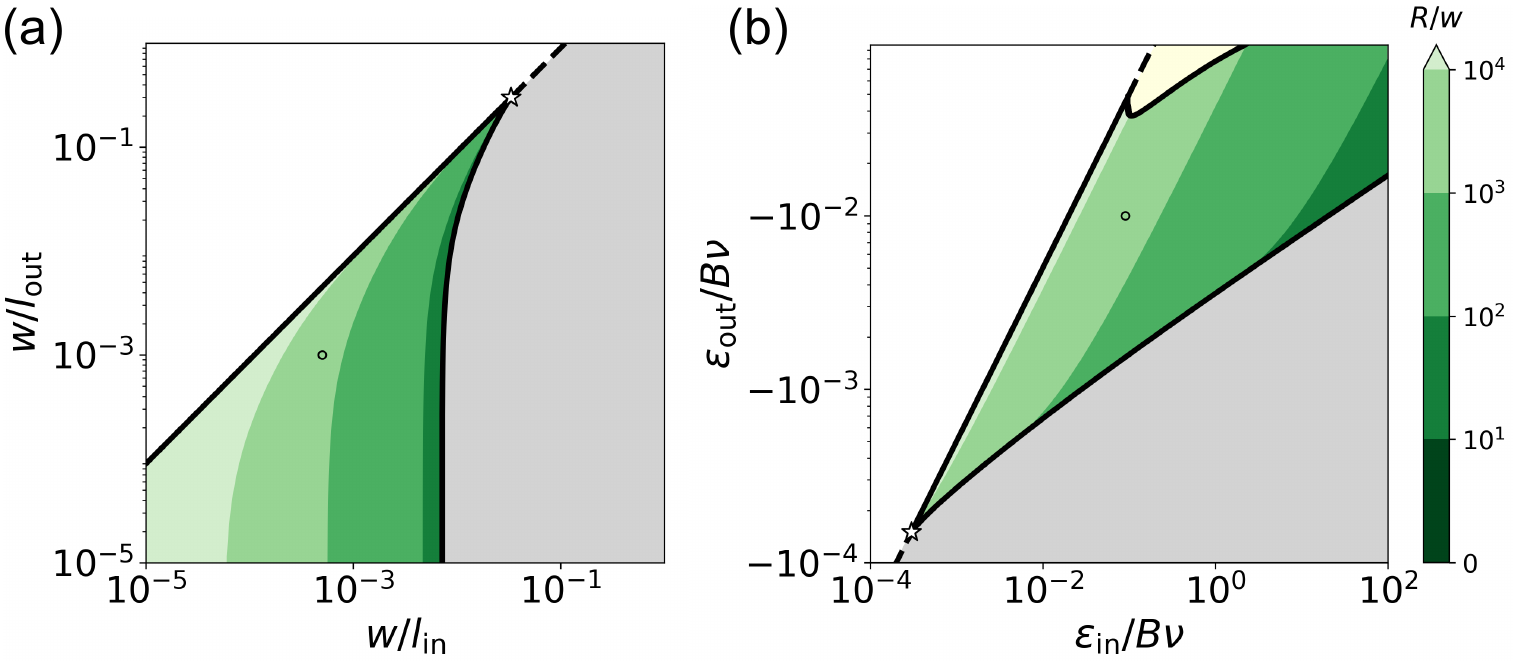}
	}
    \caption{
    \textbf{Parameter regions leading to small droplets.}
    (a) Steady state radius $R$ as a function of the two reaction-diffusion lengths, $l_\mathrm{in}$ and $l_\mathrm{out}$, for  $\varepsilon_\mathrm{in}=9\cdot10^{-2}\,B\nu$ and $\varepsilon_\mathrm{out}=-10^{-2}\,B\nu$.
    (b) $R$ as a function of the two energy scales, $\varepsilon_\mathrm{in}$ and $\varepsilon_\mathrm{out}$, for $w/l_\mathrm{in}=5\cdot10^{-4}$ and $w/l_\mathrm{out}=10^{-3}$.
    (a, b)
    The system is homogeneously enriched in component A (white region), homogeneously enriched in component B (gray region), exhibits an isolated droplet whose radius is indicated by color, or exhibits complex dynamics due to an instability inside the droplet (beige region).
    Model parameters are $\sigma=0.15\,Bw$.
    }
    \label{fig:abs_phase_log}
\end{figure}

We next ask to what extent control over the qualitative behavior of droplets translates to droplet size control.
In particular, we focus on how to realize small droplets, in order to accommodate the strict geometric constraints imposed by biological cells or synthetic applications.
To do this, we investigate droplet size control by locating the stable fixed points of \Eqref{eq:Req}.
\figref{fig:abs_phase_log} shows that boosting production of droplet material (increasing $1/l_\mathrm{out}$, $|\varepsilon_\mathrm{out}|$) increases droplet size (light shades of green), whereas boosting degradation (increasing $1/l_\mathrm{in}$, $\varepsilon_\mathrm{in}$) decreases droplet size (dark shades of green).
This is qualitatively in line with our prior observations that the former promotes volume-limited droplets and the latter promotes interface-limited droplets.
The beige region in \figref{fig:abs_phase_log}(b) corresponds to the formation of unstable spherical shells through spinodal decomposition, which generally form if droplets are large~\cite{bergmann2023liquid}. 

The trends apparent from \figref{fig:abs_phase_log} suggest that the smallest droplets overall occur in a parameter regime that simultaneously minimizes $1/l_\mathrm{out}$ and $|\varepsilon_\mathrm{out}|$, and maximizes $1/l_\mathrm{in}$ and $\varepsilon_\mathrm{in}$. 
Note, however, that this is under the condition that a stable droplet forms in the first place, i.e., that production and degradation of droplet material balance globally.
This constrains the admissible parameter values, as for a given combination of energy scales the reaction rates can only be made so fast before one dominates and destroys the stationary state; see the critical point given by \Eqref{eq:critical point}.
Concretely, this means that by maximizing the internal energy scale $\varepsilon_\mathrm{in}$ (boosting degradation), we make it possible for stable droplets to exist for increasingly high external reaction rates $1/l_\mathrm{out}$ (boosting production).
Consequently, it is possible to simultaneously maximize $\varepsilon_\mathrm{in}$ and minimize $1/l_\mathrm{out}$.
In contrast, if we minimize the external energy scale $|\varepsilon_\mathrm{out}|$ (boosting production), this significantly limits how high the internal reaction rate $1/l_\mathrm{in}$ (boosting degradation) can be.
As a result, it is generally \textit{not} possible to simultaneously minimize $|\varepsilon_\mathrm{out}|$ and maximize $1/l_\mathrm{in}$.
To obtain the smallest droplets, a tradeoff must thus be made between the relative importance of the energy scale $|\varepsilon_\mathrm{out}|$ and the reaction rate $1/l_\mathrm{in}$.

\begin{figure}
	\centering{
            \includegraphics[width=0.475\textwidth]{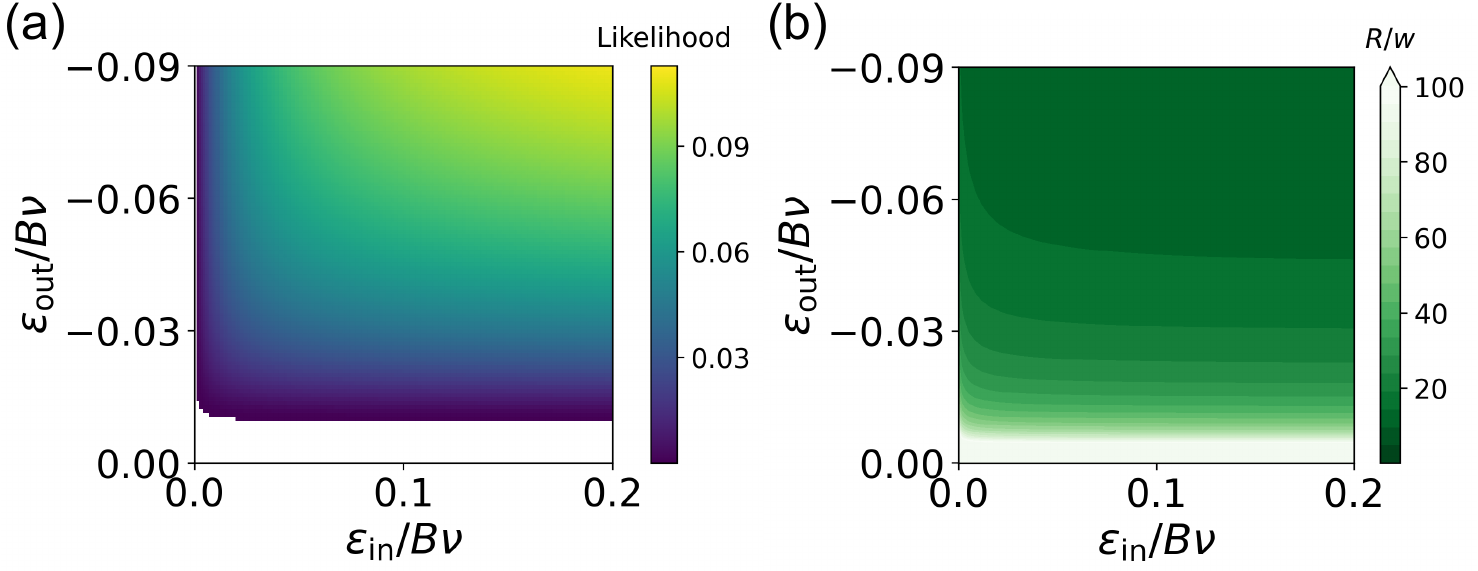}
	}
    \caption{\textbf{Parameter optimization for small droplet size}.
    (a) Likelihood of forming droplets for which $R\leq50w$ in \figref{fig:abs_phase_log}(a) as we vary $\varepsilon_\mathrm{in}/B\nu$ and $\varepsilon_\mathrm{out}/B\nu$. We weigh each order of magnitude equally and keep the scanning window constant: $w/l_\mathrm{in}\in[10^{-5},1]$, $w/l_\mathrm{out}\in[10^{-5},1]$. (b) Corresponding  achievable minimum droplet size $R/w$.
    Model parameters are $\sigma=0.15\,Bw$.
    }\label{fig:surface_area}
\end{figure}

To quantify under which conditions ``small" droplets are most likely, we compute the corresponding range of admissible reaction-diffusion lengths~$l_\mathrm{in/out}$.
To do this, we consider the area in \figref{fig:abs_phase_log}(a) that corresponds to $R\leq50w$, weighing each order of magnitude of $l_\mathrm{in/out}$ equally. %
In essence, this corresponds to the probability of realizing a ``small" droplet if one logarithmically samples all reaction-diffusion lengths plotted; the same qualitative behavior holds if we instead sample linearly. 
\figref{fig:surface_area}(a) shows that increasing either $\varepsilon_\mathrm{in}$ or $|\varepsilon_\mathrm{out}|$ increases the likelihood of realizing ``small" droplets. 
In a similar vein, \figref{fig:surface_area}(b) shows the smallest droplet size achievable for any given combination of energy scales, thus illustrating the range of possible droplet sizes. 
This figure shows that the smallest droplets are achieved by maximizing both $\varepsilon_\mathrm{in}$ and $|\varepsilon_\mathrm{out}|$.
These results suggest that maximizing the internal reaction rate, related to $1/l_\mathrm{in}$, weighs more heavily than minimizing the external energy scale, $|\varepsilon_\mathrm{out}|$, in the pursuit of small droplets.

Notably, \figref{fig:surface_area} contrasts with our prior observations that simultaneously maximizing $\varepsilon_\mathrm{in}$ and minimizing $|\varepsilon_\mathrm{out}|$ promotes volume-limited droplets ($R/l_\mathrm{in}\ll 1$).
This is because for a droplet to be volume limited, its size must be small \textit{relative} to the internal reaction-diffusion length $l_\mathrm{in}$.
As a result, although maximizing the internal reaction rate $1/l_\mathrm{in}$ produces the smallest droplets overall, it also biases droplets toward the interface-limited regime by decreasing $l_\mathrm{in}$.
Our results indicate that this shifts the balance between $|\varepsilon_\mathrm{out}|$ and $1/l_\mathrm{in}$: in the pursuit of volume-limited droplets, minimizing $|\varepsilon_\mathrm{out}|$ weighs more heavily than maximizing $1/l_\mathrm{in}$ (\figref{fig:rel_phase_log}).

Translated to the specific case discussed in the previous section---passive production of droplet material outside the droplet and active degradation inside---we propose the following design guidelines:
Increasing the passive reaction rate constant $\LambdaP$ promotes large droplets, whereas increasing the active reaction rate constant $\LambdaA$ or the chemical drive $\Delta\mu$ promotes small droplets. 
This aligns with the design guidelines for obtaining volume-limited droplets.
In contrast, while increasing $\varepsilon_0$ increases droplet size at fixed conditions, the smallest droplets overall occur at intermediate values of $0<\varepsilon_0<\Delta\mu\Delta\varphi$, due to the accessible region of the phase diagram shrinking. 
This differs from the parameter regime for which the most strongly volume-limited droplets are expected, which occurs for minimal values of $\varepsilon_0$.
We thus conclude that if one keeps the energy scales of the chemical reactions constant, increasing the rate constant of the droplet-material-producing reaction promotes larger droplets, whereas increasing the rate constant of the droplet-material-degrading reaction promotes smaller droplets.
Similarly, if one keeps the rate constants constant, increasing the energy scale driving the droplet-material-producing reaction promotes larger droplets, whereas increasing the energy scale driving of the droplet-material-degrading reaction promotes smaller droplets.
If we vary all parameters simultaneously, however, the smallest droplets are achieved by increasing both energy scales, because this choice enlarges the region of the state diagram that admits steady-state droplets.
This indicates that the design guidelines for obtaining small droplets are different from those for obtaining volume-limited droplets, reinforcing the importance of their distinction.

\section{Conclusion and discussion}

We have studied droplet stability and size control within a binary fluid model driven out of equilibrium by chemical reactions. 
We have identified four key parameters that predict the steady state of such droplets in good agreement with continuum simulations: the effective reaction-diffusion lengths, $l_\mathrm{in}$ and $l_\mathrm{out}$, and the effective energy scales, $\varepsilon_\mathrm{in}$ and $\varepsilon_\mathrm{out}$. 
We show that if one takes $l_\mathrm{in}$ as the natural length scale for droplet size, unambiguous classes of volume-limited ($R/l_\mathrm{in}\ll1$) and interface-limited ($R/l_\mathrm{in}\gg1$) droplets arise, which obey simple scaling laws. 
By tuning the relative importance of the production $(l_\mathrm{out}^{-1}, \varepsilon_\mathrm{out})$ and degradation $(l_\mathrm{in}^{-1}, \varepsilon_\mathrm{in})$ of droplet material, we can transition between these regimes.
Although this view of size control governs qualitative droplet behavior, it does not necessarily accommodate exact geometric constraints, as are present in biological cells and synthetic applications.
Importantly, we show that the design guidelines for realizing small droplets deviate qualitatively from the design guidelines for realizing volume-limited droplets ($R/l_\mathrm{in}\ll1$).

Concretely, our findings indicate that in order to achieve ``small" droplets, a tradeoff must be made between the energetics and the kinetics of the associated chemical reactions.
This is because the reaction rates dictate the range of energy scales that admit steady-state droplets and \textit{vice versa}, implying that it is not generally possible to optimize for both simultaneously.
If we take ``small" to mean volume limited, it turns out that we must prioritize the energetics, and our recommendation is to minimize $l_\mathrm{out}^{-1}$, $|\varepsilon_\mathrm{out}|$ and maximize $l_\mathrm{in}^{-1}$, $\varepsilon_\mathrm{in}$.
If we instead take ``small" to refer to the absolute size of the droplet, we must prioritize the kinetics, resulting in the recommendation to minimize $l_\mathrm{out}^{-1}$ and maximize $l_\mathrm{in}^{-1}$, $\varepsilon_\mathrm{in}$, $|\varepsilon_\mathrm{out}|$.
This tradeoff suggests that the qualitative behavior of the droplet and its absolute size could in principle be tuned separately.
We speculate that this might allow cells and synthetic applications to further specialize condensates to fulfill specific functions, such as buffering or regulation \cite{klosin2020phase,deviri2021physical,zechner2025concentration,Monchaux2025buffering,snead2019control,soding2020mechanisms,jeon2025emerging}, even under geometric constraints.

More broadly, our work serves as a point of departure for understanding droplet size control in more general contexts.
An obvious extension would be to study emulsions of closely interacting droplets \cite{kostler2025advection,bauermann2024critical}, rather than isolated droplets.
Similarly, it would be interesting to study systems with more than two components \cite{qiang2024multicomponent,bauermann2024critical}, which exhibit conserved quantities that can vary in space.
In addition, droplets in many biological contexts contend with much more complex, elastic micro-environments, which are known to affect their dynamics \cite{vidal2020theory,vidal2021cavitation,paulin2025elastic,Ronceray2022}.
Finally, we considered droplet size control in the deterministic limit, but how thermal fluctuations affect droplet size control remains an outstanding question; especially considering that thermal energy should be relevant on the typical length and energy scales of many (biological) droplets.

\subsection*{Acknowledgments}
We gratefully acknowledge funding from the Max Planck Society and the European Research Council (ERC, EmulSim, 101044662).

\bibliography{Bibliography.bib}

\end{document}